\documentclass[oneside]{amsart}

\newcommand{\R}{\mathds R}
\newcommand{\I}{\mathds I}

\usepackage{epsfig}
\usepackage{times}
\usepackage{dsfont}
\usepackage{amsmath}
\usepackage[hypertex]{hyperref}

\theoremstyle{plain}\newtheorem{teo}{Theorem}[section]
\theoremstyle{plain}\newtheorem{lem}[teo]{Lemma}
\theoremstyle{plain}\newtheorem{prop}[teo]{Proposition}
\theoremstyle{plain}
\theoremstyle{definition}
\theoremstyle{remark}\newtheorem{rem}[teo]{Remark}
\theoremstyle{remark}\newtheorem{ex}[teo]{Example}
\theoremstyle{plain}

\numberwithin{equation}{section}

\title[ Collapse of homogeneous scalar fields ]%
{Genericity of blackhole formation in the gravitational collapse of
homogeneous self-interacting scalar fields}

\author[R.\ Giamb\`o ,\ F.\ Giannoni]{Roberto Giamb\`o, Fabio Giannoni}
\address{Dipartimento di Matematica e Informatica \hfill\break\indent
Universit\`a di Camerino\hfill\break\indent Italy}
\email{roberto.giambo@unicam.it, fabio.giannoni@unicam.it}

\author[G.\ Magli]{Giulio Magli}
\address{Dipartimento di Matematica \hfill\break\indent Politecnico di
Milano \hfill\break\indent Italy} \email{giulio.magli@polimi.it}

\begin{document}
\begin{abstract}

The gravitational collapse of a wide class of self-interacting
homogeneous scalar fields models is analyzed. The class is
characterized by certain general conditions on the scalar field
potential, which, in particular, include both asymptotically
polynomial and exponential behaviors. Within this class, we show
that the generic evolution is always divergent in a finite time, and
then make use of this result to construct radiating star models of
the Vaidya type. It turns out that blackholes are generically formed
in such models.

\end{abstract}

\maketitle

\section{Introduction}\label{sec:intro}


Scalar fields attracted a great deal of attention in cosmology and
in relativistic astrophysics in the last twenty years. It is,
indeed, worth mentioning the fundamental role that they play in
modeling the early universe and, on the other hand, the relevance
of the self-gravitating scalar field model as a test-bed for the
Cosmic Censorship conjecture, that is, whether the singularities
which form at the end state of the gravitational collapse are always
covered by an event horizon, or not \cite{ns,j-rev}. In particular,
in an important series of papers, Christodolou studied massless
scalar field sources minimally coupled to gravity. For this very
special case the formation of some naked singularities has been
shown, but it has been also shown that such singularities are non
generic with respect to the choice of the initial data \cite{c1,c2}.

The massless scalar field is a useful model in studies of
gravitational collapse since its evolution equation in absence of
gravity - the massless Klein-Gordon equation on Minkowski spacetime
- is free of singular solutions. However, of course,  the general
case in which the field has non-zero self-interaction potential
(which of course include the massive scalar field as a quadratic
potential) is worth inspecting from the theoretical point of view.
Further to this, in recent years, interest in models with non
vanishing potentials aroused from string theory, since, for
instance, dimensional reduction of fundamental theories to four
dimensions typically gives rise to self-interacting scalar fields
with exponential potentials, coupled to four-dimensional gravity. In
this context, the case of potentials which have a negative lower
bound plays a special role, since such potentials generate the anti
De Sitter space as equilibrium solutions \cite{hor,hor1}.

As a consequence of their "fundamental" origin, exponential
potentials have been widely considered in the
cosmologically-oriented literature (see \cite{russo} and references
therein) also with the aim of uncovering possible large-scale
observable effects \cite{fosh,st1,st2,topo}. A few examples of non
trivial potentials (to be discussed in full details in the sequel)
have been studied in the gravitational collapse scenario as well. In
these papers formation of naked singularities has been found
\cite{hsf,joshi}.

So motivated, in the present paper  we study the dynamical behavior
of FRW (spatially flat)  scalar field models, imposing only very
general conditions on the scalar field potential; our hypotheses
essentially reduce to ask the potential to be bounded from below and
divergent when the field diverges. The evolution of collapsing
solutions or, equivalently, time backwards expanding solutions, has
been studied previously in the literature mainly making use of
dynamical system approaches. From this point of view Foster's work
is worth noticing; indeed, in \cite{fos} existence of past
singularities of expanding, spatially  flat cosmologies are found in
terms of sufficient conditions on the (nonnegative) potential
$V(\phi)$ -- in particular, the proof relies on the existence of a
coordinate transformation that brings $[\phi_0,+\infty[$ in
$]0,\epsilon]$ , and satisfies some other regularity conditions
together with the transformed potential (see \cite[Definition 2]{fos}).
In these cases, the behavior is proved to be essentially similar to
the behavior in the massless case, and so a past singularity exists
and it is stable with respect to the choice of the data.

Other worth-mentioning works are those related to the case of
non-minimal coupling, such as
 \cite{bojo}, in which  flat FRW model are studied using conformal
transformations that bring the models into the standard form, and
\cite{vc1}, where singularity formation for non minimal coupling
models is analyzed for some specific form of the potential.

As mentioned above, in this paper we consider a class of potentials
$\mathfrak C$ which is quite more general with respect to those
previously considered. In particular, we weaken assumptions about
the behavior of the potential at infinity (which are necessary in
the dynamical systems approach) thanks to an accurate use of the
Implicit Function Theorem. Indeed a suitable estimate of the
involved neighborhoods is used to obtain a result of existence,
uniqueness, and instability of solution for a particular kind of
ordinary differential equation of first order to which the standard
theory does not apply straightforwardly (Theorem \ref{thm:ode}).
Another key difference with previously published results is that the
potentials may assume negative values. Non negativity of the
potential means that the total energy of the scalar field has a
positive lower bound (using this constraint many results for scalar
field cosmologies have been actually found
\cite{fos,mir2,mir3,r1,r2,rvc3}; see also \cite{exp}); however, in
many issues and especially when string theory comes into play, it
becomes relevant to inspect spacetimes in which the potential still
has a lower bound but this bound is negative; under certain
conditions indeed, potentials of this kind generate solutions with
positive total energy \cite{hert}. Since, in particular , a constant
negative potential generates an "equilibrium" state which is just
the Anti de Sitter solution, within this framework models which are
asymptotically anti-De Sitter can be inspected \cite{exp,hor,hor1}.

Within the class $\mathfrak C$ of potentials (to be defined in
precise technical terms in the section below) we prove that - except
for a zero measure set of initial data - all the solutions become
singular in a finite time. Considering such solutions as sources for
models of collapsing objects (with generalized Vaidya solutions as
the exterior) we study the nature - blackhole or naked singularity -
of such objects. It turns out that, except possibly for a
zero-measure set of initial data, all these "scalar field stars"
collapse to black holes. It follows that the existing examples of
naked singularities which occur in the literature must be
non-generic within the set of possible solutions, and indeed we show
in details how this occurs.

Homogeneous scalar field models with self-interacting potentials
thus respect the weak cosmic censorship hypothesis, at least for a
wide class of potentials. Of course, however, the problem of
extending this result to the non-homogeneous case remains open.

\section{Collapsing models}\label{sec:cosmo}

We are going to consider here homogeneous, spatially flat
spacetimes
\begin{equation}\label{eq:g}
g=-\mathrm dt\otimes\mathrm dt+a^2(t)\left[\mathrm dx^1\otimes\mathrm
dx^1+\mathrm dx^2\otimes\mathrm dx^2+\mathrm dx^3\otimes\mathrm dx^3\right]
\end{equation}
where gravity is coupled to a scalar field $\phi$, self-interacting
with a scalar potential $V(\phi)$. Assigning this function is
equivalent to assign the physical properties of the matter source,
and thus the choice of the potential can be seen as a choice of the
equation of state for the matter model. Therefore, as occurs in
models with "ordinary" matter (e.g., perfect fluids), the choice of
the "equation of state" must be restricted by physical
considerations. Stability of energy, in particular, requires the
potential to have a lower bound. We shall thus consider here
potentials of class $C^2(\R)$ bounded from below; the lower bound
of the potential need not be positive however, so that class of
solutions which include AdS minima can be considered. Further, we
need a set of technical assumptions.

We say that $V$ belongs to the set $\mathfrak C$ if the following
conditions are satisfied:

A1) (Structure of the set of critical points) The critical points of
$V$ are isolated and they are either minimum points or
non-degenerate maximum points

A2) (Existence of a suitable bounded "sub-level" set) There exists $\alpha < \beta \in \mathds{R}$
such that
\[
\phi\ge\beta\Longrightarrow V'(\phi)>0,\, \phi\le\alpha\Longrightarrow V'(\phi)<0.
\]
Moreover, setting $V_* = \max_{[\alpha,\beta]}V$, it is
\[
\lim_{\phi\rightarrow -\infty}V(\phi) > V_{*},\,\lim_{\phi\rightarrow +\infty}V(\phi) > V_{*}.
\]

A3) (Growth condition) The function defined in terms of $V$ and its
first derivative $V'$ as
\begin{equation}\label{eq:Y}
u(\phi):=\frac{V'(\phi)}{2\sqrt 3V(\phi)},
\end{equation}
satisfies to
\begin{align}
&\limsup_{\phi\to\pm\infty} |u(\phi)|< 1, \label{eq:u-infty}\\
&\exists\lim_{\phi\to\pm\infty} {u'(\phi)} \,
(=0).\label{eq:uprime-infty}
\end{align}

We stress that the set $\mathfrak C$ defined in terms of the above
conditions is sufficiently large to contain all the potentials which
have been considered in physical applications so far. For instance,
it contains the standard quartic potential ($V(\phi)=\frac12
m^2\phi^2+\lambda^2\phi^4$) and, more generally, all the potentials
bounded from below whose asymptotic behavior is polynomial (i.e.
$\phi^{2n}$ for $|\phi|\to\infty$, $n \geq1$).

\bigskip

Once a potential has been chosen, the Einstein field equations
allow, in principle, to obtain the dynamics of the coupled system
composed by the field and the metric. These equations are given by
\begin{subequations}
\begin{align}
&(G^0_0=8\pi T^0_0):\qquad-\frac{3\dot
a^2}{a^2}=-(\dot\phi^2+2V(\phi)),\label{eq:original-efe1}\\
&(G^1_1=8\pi T^1_1):\qquad-\frac{\dot a^2+2a\ddot
a}{a^2}=(\dot\phi^2-2V(\phi)),\label{eq:original-efe2}
\end{align}
\end{subequations}
the Bianchi identity is of course, due to the Noether theorem,
equivalent to the equation of motion for the field:
\begin{equation}\label{eq:Bianchi}
T^\mu_{\,\,0;\mu}=-2\dot\phi\left(\ddot\phi+V'(\phi)+3\frac{\dot
a}a\dot\phi\right)=0.
\end{equation}
Denoting the energy density of the scalar field by
\begin{equation}\label{eq:energy}
\epsilon=\dot\phi^2+2V(\phi),
\end{equation}
We will actually consider the following system:
\begin{subequations}
\begin{align}
&\left(\frac{\dot a}a\right)^2={\frac\epsilon 3}.\label{eq:initialefe2}\\
&\ddot\phi+V'(\phi)=-3\frac{\dot a}a\,\dot\phi,\label{eq:initialefe1}
\end{align}
\end{subequations}

\bigskip

Of course, it is mandatory to extract the square root in the first
equation. Since we are considering the collapsing model we choose
$\dot a$ to be negative and therefore our final system is composed
by
\begin{subequations}
\begin{align}
&\frac{\dot a}a=-\sqrt{\frac\epsilon 3}.\label{eq:efe2}\\
&\ddot\phi+V'(\phi)=\sqrt
3\sqrt\epsilon\,\dot\phi.\label{eq:efe1}
\end{align}
\end{subequations}
Let us observe that, using \eqref{eq:energy} and \eqref{eq:efe1},
the following relation can be seen to hold:
\begin{equation}\label{eq:3}
\dot\epsilon=2\sqrt 3\sqrt\epsilon\,\,\dot\phi^2,
\end{equation}
that is $\epsilon(t)$ is monotone. Observe that the two cases - expansion and
collapse - are connected by the fact that, in the expanding case,
there might be the possibility of reaching a vanishing value of
$\epsilon$ in a finite time $t_0$. If this occurs, the model will be
ruled from $t_0$ onward by the equations for the collapsing
situation. Actually, it will be proved that all such solutions are
divergent at some finite time for almost every choice of the initial
data, thus yielding a singularity as well.

In what follows, we are going to focus on solutions of the equations
\eqref{eq:initialefe2}--\eqref{eq:initialefe1} -- or, better,
\eqref{eq:efe2}--\eqref{eq:efe1}. Actually, it can be proved
\cite{jmp} that, if $a\ne 0$ for all $t \geq 0$ and $(a,\phi)$ are
$\mathcal C^2$ functions that solve
\eqref{eq:original-efe1}--\eqref{eq:original-efe2} (with a
non-constant $\phi(t)$) then they are solutions of
\eqref{eq:initialefe2}--\eqref{eq:initialefe1} as well. To prove the
converse, it suffices to derive \eqref{eq:initialefe2} w.r.t. time
and use \eqref{eq:initialefe1} to deduce that if (in some interval
$[a,b]$), $\dot a$ is not identically zero -- that is, $\epsilon\ne
0$ -- then solutions of
\eqref{eq:initialefe2}--\eqref{eq:initialefe1} are also solutions of
\eqref{eq:original-efe1}--\eqref{eq:original-efe2}. Finally, the
unique case in which $\epsilon(t)=0,\forall t>0$ and the Einstein
Field Equations are satisfied is given by the trivial solution
$a(t)=a_0$, $\phi(t)=\phi_0$, with $V(\phi_0)=0$.

\section{Qualitative analysis of the collapsing models}\label{sec:collapse}

The aim of the present section is to study the qualitative behavior
of the solutions of \eqref{eq:efe2}--\eqref{eq:efe1} whenever $V\in
\mathfrak C$. The initial data (on $t=0$, say) for the system can be
fully characterized in terms of the values of the scalar field
$\phi_0=\phi(0)$ and of its first derivative. In what follows we
will be mainly interested in constructing models of collapsing
object and therefore we consider only solutions which satisfy the
weak energy condition on the initial data surface. Therefore, we
assume
\begin{equation}\label{eq:wec}
\epsilon_0=\epsilon(0)=\dot\phi^2(0)+2V(\phi(0))\ge 0.
\end{equation}
Given a solution $\phi(t)$, let $\I\subseteq[0,+\infty)$ be the
maximal right neighborhood of $t=0$ where the solution is defined,
and call $t_s=\sup\I$. The main result of this section is that the
solution $\phi(t)$ diverges in a finite time for almost every choice
of the initial data satisfying \eqref{eq:wec}, that is $t_s$ is
finite for almost all solutions. This result will in turn allow us
(next section) to construct models of collapsing objects by matching
these solutions with suitable Vaidya metrics.

To proceed further, we start by considering only data such that
$\epsilon_0$ is "sufficiently large", more precisely
\begin{equation}\label{eq:initial}
\epsilon_0 > 2V^*,
\end{equation}
where, for each choice of the potential $V$ in $\mathfrak C$, the constant
$V^*$ is the chosen as the same which appears in (A2). Once the
result will be proven for these data, we will extend its validity by
removing the restriction \eqref{eq:initial}.

We start our study with the following lemma.

\begin{lem}\label{thm:phidotphi}
If $\phi(t)$ is bounded, also $\dot\phi(t)$ is bounded, and
$t_s=+\infty$.
\end{lem}
\begin{proof}
Let $K$ a bounded set such that $\phi(t)\in K,\,\forall t\in \I$,
and argue by contradiction, assuming $\dot\phi$ not bounded, and
supposing that $\sup_\I\dot\phi(t)=+\infty$ -- the same argument can
be used if $\dot\phi$ is  unbounded only below. Let $\bar t$ such
that
\[
\dot\phi(\bar t)>\frac{\sup_{K}V'(\phi)}{\sqrt{3\epsilon_0}},
\]
therefore \eqref{eq:efe1} implies that $\ddot\phi(\bar t)>0$, and
this means that $\dot\phi$ is increasing in $\bar t$, so
\eqref{eq:efe1} shows that $\ddot\phi\ge 0,\,\forall t\geq\bar t$,
and then $\dot\phi$ is eventually increasing, namely $\lim_{t\to
t_s}\dot\phi(t)=+\infty$. This shows that $t_s<+\infty$, otherwise
it would be $\phi(t)-\phi_0=\int_0^t\dot\phi(\tau)\,\mathrm d\tau$,
which would diverge as $t\to+\infty$. Then $t_s\in\R$.

With the position
\[
\lambda(t)=\frac1{\dot\phi(t)},
\]
equation \eqref{eq:efe1} implies
\[
\dot\lambda(t)=\frac{V'(\phi(t))}{\dot\phi(t)^2}-\sqrt
3\sqrt{1+\frac{2V(\phi(t))}{\dot\phi(t)^2}}\xrightarrow{t\to t_s^-}-\sqrt 3,
\]
and then $\lambda(t)$, near $t=t_s$, behaves like $\sqrt 3(t_s-t)$.
Since $\dot\phi(t)$ positively diverges,  $\lim_{t\to
t_s}\phi(t)=\phi^*$ exists, and is finite since $\phi(t)$ is bounded
by hypothesis. But the quantity
\[
\phi(t)-\phi^*=\int_{t_s}^t\dot\phi(\tau)\,\mathrm
d\tau=\int_{t_s}^t\frac1{\lambda(\tau)}\,\mathrm d\tau
\]
diverges, which is absurd. This shows that $\dot\phi(t)$ must be
bounded too.

To prove that $t_s=+\infty$, we proceed again by contradiction. Let
$t_k\to t_s$ be a sequence such that the sequence
$(\phi(t_k),\dot\phi(t_k))$ converges to a finite limit
$(\phi_*,\dot\phi_*)$. Solving the Cauchy problem \eqref{eq:efe1}
with initial data $(\phi_*,\dot\phi_*)$ shows that the solution
$\phi(t)$ is $\mathcal C^1$, the solution could be prolonged on a
right neighborhood of $t_s$.
\end{proof}

\begin{prop}\label{thm:limsup}
The function $\phi(t)$ is unbounded.
\end{prop}
\begin{proof}
Let by contradiction $\phi(t)$ be bounded. Then, by Lemma
\ref{thm:phidotphi}, $\vert\dot\phi(t)\vert\le M$ for a suitable
constant $M$, and $t_s=\sup\I=+\infty$. In particular $V(\phi(t)),
V'(\phi(t)), \epsilon(t),\ddot\phi(t)$ are bounded also. Then Lemma
\ref{lem:dotphi0} in Appendix \ref{sec:eps}
 says that $\lim_{t\to+\infty}\dot\phi(t)=0$, that implies
$\lim_{t\to+\infty}2V(\phi(t))=\lim_{t\to+\infty}\epsilon(t)\ge\epsilon_0
> 2V_*$, and so $V(\phi(t)) > V_{*}$ for any $t$ sufficiently large.
Then, by assumption A2), for large $t$, $\phi(t)$ moves
outside the interval $[\alpha,\beta]$ described in assumption A2), namely
in a region where $V$ is
invertible. Since $V(\phi(t))$ converges this happen also for
$\phi(t)$ which converges to a point $\phi_* \not \in ]\alpha,\beta[$. Then by \eqref{eq:efe1}, $\ddot\phi(t)\to-V'(\phi_*)\ne 0$
(since $\dot \phi (t) \rightarrow 0$), that is absurd. Then
$\limsup_{t\to t_s}\vert\phi(t)\vert=+\infty$.
\end{proof}

We have shown that $\phi(t)$ is not bounded. Now we want to show
that, actually, $\phi(t)$ monotonically diverges in the approach to
$t_s$.

\begin{rem}\label{rem:rho}
We observe that the quantity
\begin{equation}\label{eq:rho}
\rho(t):=\frac{2V(\phi(t))}{\dot\phi(t)^2},
\end{equation}
satisfies the equation
\begin{equation}\label{eq:ODErho}
\dot\rho=2\sqrt3\dot\phi\,\rho\sqrt{(1+\rho)}\left(u(\phi)\sqrt{1+\rho}-\mathrm{sgn}\,(\dot\phi)\right).
\end{equation}
\end{rem}

\begin{prop}\label{thm:div}
Let $V\in\mathfrak C$, and let $\phi:\I\subseteq\R\to\R$ the
solution of \eqref{eq:efe1}. Then $\dot\phi(t)$ is eventually nonzero, and
\[
\lim_{t\to\sup\I}|\phi(t)|=+\infty.
\]
\end{prop}
\begin{proof}
Note that $\dot \phi(t) = 0$ imply $V(\phi(t)) = \frac12\epsilon(t) \geq \frac12\epsilon_{0} > V_{*}$,
so $\phi \not \in [\alpha,\beta]$. Moreover if $\phi(t) > \beta$ , it is $V'(\phi(t)) > 0$ so $\ddot \phi(t) < 0$,
while if $\phi(t) < \alpha$ it is $\ddot \phi(t) > 0$, Therefore if
by contradiction, $\dot \phi(t) = 0$ for infinitely many values of $t$, we see that $\phi(t)$ crosses the interval
$[\alpha,\beta]$ infinitely many times.

Then there exist sequences $\sigma_k < \tau_k$ going to $t_s$ such that
\[
\phi(\sigma_k)=\alpha,\qquad\phi(\tau_k)=\beta,\qquad\phi([\sigma_k\tau_k])\subseteq[\alpha,\beta],
\]
and
\[
\beta - \alpha =\phi(\tau_k)-\phi(\sigma_k)=\int_{\sigma_k}^{\tau_k}\dot\phi(t)\mathrm
dt\le M(\tau_k-\sigma_k),
\]
and so $\tau_k-\sigma_k\ge \tfrac {\beta-\alpha}M$. Moreover, since $\phi([\sigma_k,\tau_k]) \subseteq [\alpha,\beta]$ it is
\[
\epsilon_{0}\leq\epsilon(t) = (\dot \phi(t))^{2} +2V(\phi(t)) \leq (\dot \phi(t))^{2} +2V_{*}
\]
for any $t \in [\sigma_k,\tau_k]$. Then
\[
(\dot \phi(t))^{2} \geq \epsilon_{0} - 2V_{*} > 0
\]
for any $t \in [\sigma_k,\tau_k]$.

Therefore,  by \eqref{eq:3},
\[
\sqrt{\epsilon(\tau_k)}-\sqrt{\epsilon(\sigma_k)}=\int_{\sigma_k}^{\tau_k}
\frac{\dot\epsilon}{2\sqrt\epsilon}\,\mathrm dt=
\sqrt 3\int_{\sigma_k}^{\tau_k}\dot\phi^2\,\mathrm dt\ge(\epsilon_{0} -2V_{*})\sqrt{3}
(\tau_k-\sigma_k)
\ge(\epsilon_{0} -2V_{*})\sqrt{3}\tfrac{\beta-\alpha}M,
\]
and then $\epsilon(t)$ must diverge.

Now, recalling (\ref{eq:u-infty}), let $\theta>0$ be a sufficiently
small constant, and let $r_k\to t_s$ a sequence such that
$\phi(r_k)=\bar\phi$, with $u(\phi) \leq 1-\theta$ as
$\phi\ge\bar\phi$. Then
$\dot\phi(r_k)^2=\epsilon(r_k)-2V(\bar\phi)\to+\infty$. Let us
suppose that $\dot\phi(r_k)$ is unbounded above -- analogously one
can argue if it is only unbounded below. Up to subsequences, we can
suppose that $\dot\phi(r_k)$ positively diverges. Moreover,
recalling \eqref{eq:rho}, $\rho(r_k)\to 0$ and then, if $k$ is
sufficiently large,
\[
\vert\rho(r_k)\vert<\left(\frac1{1-\theta}\right)^2-1,
\]
so that, since $\phi(r_k)=\bar\phi$ and $u(\bar\phi) \leq 1-\theta$,
we have
\[
u(\phi(r_k))\sqrt{1+\rho(r_k)}-1<0.
\]
Then, $\dot\rho(r_k)<0$, that is $\rho(t)$ is decreasing at $r_k$.
But we can observe that $\phi(t)$ is increasing in $r_k$, ensuring
$u(\phi(t))<1-\theta$, and so, in a right neighborhood of $r_k$,
\[
u(\phi(t))\sqrt{1+\rho(t)}-1<(1-\theta)\sqrt{1+\rho(t)}-1
\]
that is decreasing. We conclude that the function $\rho(t) = \tfrac{2V(\phi(t))}{\dot \phi^2(t)}$
decreases for $t>r_k$, until $t$ equals a local maximum $t_k$, where
$\dot\phi$ vanishes.
Then there exists $\kappa>0$ such that $2V(\phi(t))\le\kappa\dot\phi(t)^2$
until $t$ equals a local maximum $t_k$. But this would imply $\epsilon(\phi(t_k))=0$,
getting
a contradiction. Then $\dot \phi$ is eventually non zero. This
implies that there exists $\lim\limits_{t\rightarrow
+\infty}|\phi(t)|$ so, by Proposition \ref{thm:limsup}, it is
$+\infty$.
\end{proof}

We have shown so far that $\phi(t)$ diverges, as $t$ approaches
$t_s=\sup\I$. In the following, we will show that $t_s\in\R$ for
almost every solution, in the sense that there exists a set of
initial data, whose complement has zero Lebesgue measure, such that the
solution is defined until a certain finite comoving time $t_s$
(depending on the data).

Henceforth we will suppose (just to fix ideas) $\phi(t)$ \emph{positively} diverging and $\dot \phi(t) > 0$
$\forall t\ge\bar t$. Then, in the interval $[\bar t,+\infty)$,
$\rho$ can be seen as a function of $\phi$, that satisfies by
\eqref{eq:ODErho} the ODE
\begin{equation}\label{eq:eva}
\frac{\mathrm d\rho}{\mathrm
d\phi}=2\rho\sqrt{3(1+\rho)}\left(u(\phi)\sqrt{1+\rho}-1\right).
\end{equation}
Notice that considering $\phi$ as positively diverging is not
restrictive, since it can be easily shown that $\rho(-\phi)$
satisfies the same equation as $\rho$.

We now state the following crucial result:

\begin{lem}\label{thm:perla}
Except at most for a measure zero set of initial data, the function
$\rho(t)$
goes to zero for $t\to\sup\I$.
\end{lem}

To obtain the proof we need a result of
of existence, uniqueness, and
instability of solution for a particular kind of ordinary
differential equation of first order to which the standard theory
does not apply straightforwardly.
The proof of such a result is based on a suitable estimate of the involved neighborhoods when applying Implicit Function Theorem.

\begin{teo}\label{thm:ode}
Let us consider the Cauchy problem
$$
h(s)\dot z(s)=f(s,z(s))+g(s),\qquad z(0)=0.
$$
and $\beta$ is a positive constant such that the following
conditions hold:
\begin{enumerate}
\item $h\in{\mathcal C}^0([0,\beta],\R)$ such that $h(0)=0$, $h(s)>0$ in
$]0,\beta]$, and  $h(s)^{-1}$ is not integrable in $]0,\beta]$;
\item $g\in{\mathcal C}^0([0,\beta],\R)$ such that $g(0)=0$;
\item $f\in{\mathcal C}^{0,1}([0,\beta]\times\R,\R)$ such that $f(s,0)\equiv
0$, $\tfrac{\partial f}{\partial z}(0,0)< 0$, and
\item $\exists\rho>0$ such that $\tfrac{\partial f}{\partial z}(s,z)$ is
uniformly Lipschitz continuous with respect to $z$ in
$[0,\beta]\times[-\rho,\rho]$, that is $\exists L>0$ such that,
if $\vert z_1\vert, \vert z_2\vert\le\rho$, $s\in [0,\beta]$,
then $\vert \tfrac{\partial f}{\partial
z}(s,z_1)-\tfrac{\partial f}{\partial z}(s,z_2)\vert\le L \vert
z_1-z_2\vert$.
\end{enumerate}

Then, there exists $\alpha<\beta$ such that the above Cauchy problem
admits a unique solution $z(s)$ in $[0,\alpha]$. Moreover, this
solution is the only function solving the differential equation
$h(s)\dot z(s)=f(s,z(s))+g(s)$ with the further property
$\liminf_{s\to 0}\vert z(s)\vert=0$.

\end{teo}

\begin{proof}
Let us set $\alpha<\beta$ free to be determined later, and let
${\mathcal X}_\alpha$ be the space
$$
{\mathcal X}_\alpha=\{z\in{\mathcal C}^0([0,\alpha])\cap{\mathcal
C}^1(]0,\alpha])\,:\,z(0)=0,\,\lim_{s\to 0^+}h(s)\dot z(s)=0\}.
$$
It can be proved that ${\mathcal X}_\alpha$ is a Banach space,
endowed with the norm $\Vert z\Vert_\alpha=\Vert z\Vert_\infty +
\Vert h\dot z\Vert_\infty$.

Let also be
$$
{\mathcal Y}_\alpha=\{\lambda\in{\mathcal
C}^0([0,\alpha])\,:\,\lambda(0)=0\}
$$
a (Banach) space endowed with the $L^\infty$--norm and let us
consider the functional
$$
{\mathcal F}:{\mathcal X}_\alpha\to{\mathcal Y}_\alpha, \qquad {\mathcal
F}(z)(s)=h(z)\dot z(s)-f(s,z(s)).
$$
It is easily verified that $\mathcal F$ is a $\mathcal C^1$
functional, with tangent map at a generic element $z\in\mathcal
X_\alpha$ given by
\[
\left(\text d\mathcal F(z)[\xi]\right)(s)= h(s)\dot\xi(s)-\frac{\partial
f}{\partial z}(s,z(s))\xi(s),
\]
where $\xi\in  X_\alpha$. Observing that $g\in{\mathcal Y}_\alpha$,
we want to find $\alpha$ such that the equation
\begin{equation}\label{eq:inv}
\mathcal F(z)=g
\end{equation}
has a unique solution $z \in  X_\alpha$. To this aim, we will
exploit an Inverse Function scheme, and we will prove that $\mathcal
F$ is a local homeomorphism from a neighborhood of $z_0\equiv 0$ in
$\mathcal X_\alpha$ onto a neighborhood of $\mathcal F(z_0)\equiv 0$
in $\mathcal Y_\alpha$, that includes $g$. This will be done using
neighborhoods with radius independent of $\alpha$ and this will be
crucial to obtain the uniqueness result.

In the following we review the classic scheme (see for instance
\cite{Berg}) for reader's convenience. Let $R(z)=\mathcal
F(z)-\mathcal F(0)-\text d\mathcal F(0)[z]$; then \eqref{eq:inv} is
equivalent to find $z$ such that $R(z)+\text d\mathcal F(0)[z]=g$,
and therefore, if $d\mathcal F(0)$ is invertible, to prove the
existence of a unique fixed point of the application $T$ on
$\mathcal X_\alpha$,
\begin{equation}\label{eq:T}
T(z)=(d\mathcal F(0))^{-1}[g-R(0,z)].
\end{equation}
We first show that $T$ is a contraction map from the ball
$B(0,\delta)\subseteq\mathcal X_\alpha$ in itself, provided that
$\delta$ and $\Vert g\Vert_\infty$ are sufficiently small
(independently by $\alpha$). The following facts must be proven to
this aim:
\begin{enumerate}
\item there exists a constant $M$, independent on $\alpha$,
such that, $\forall z_1,z_2\in\mathcal X_\alpha$ with $\Vert
z_1\Vert_\alpha,\Vert z_2\Vert_\alpha\le 1$, it is
\begin{equation}\label{eq:F-lip}
\Vert\text d\mathcal F(z_1)-\text d\mathcal F(z_2)\Vert
\le M \Vert z_1-z_2\Vert_\alpha
\end{equation}
(the norm on the left hand side refers to the space of linear
applications from $\mathcal X_\alpha$ to $\mathcal Y_\alpha$).
\item there exists a constant $C$, independent on $\alpha$, such that
\begin{equation}\label{eq:Fprime}
\Vert \text d\mathcal F(0)^{-1}\Vert \le C
\end{equation}
(here the norm refers to the space of linear applications from
$\mathcal Y_\alpha$ to $\mathcal X_\alpha$).
\end{enumerate}
If the above facts hold, given $z_1,z_2\in \mathcal X_\alpha$, then
\begin{multline*}
\text d\mathcal F(0)[T(z_1)-T(z_2)]\\ =R(z_2)-R(z_1)=\mathcal
F(z_2)-\mathcal F(z_1)-\text d\mathcal F(0)[z_2-z_1]\\ =\int_0^1
\left(\text d\mathcal F(t z_2+(1-t)z_1)-\text d\mathcal
F(0)\right)[z_2-z_1]\,\text dt,
\end{multline*}
hence, if in addition $z_1,z_2\in B(0,\delta)$,
\begin{multline*}
\Vert T(z_1)-T(z_2)\Vert_\alpha\le \\ \Vert (\text d\mathcal
F(0))^{-1}\Vert \left(\int_0^1 \Vert \left(\text d\mathcal F(t
z_2+(1-t)z_1)-\text d\mathcal F(0)\right)\Vert \,\text
dt\right)\,\Vert z_1-z_2\Vert_\infty\le 2 M\,C\,\delta \Vert
z_1-z_2\Vert_\alpha.
\end{multline*}
and therefore:
\begin{itemize}
\item $T$ is a contraction, taking $\delta$ such that $K:=2MC\delta<1$;
\item since $\Vert T(z)\Vert_\alpha\le \Vert T(z)-T(0)\Vert_\alpha+\Vert
T(0)\Vert_\alpha\le K \Vert z\Vert_\alpha+ \Vert (d\mathcal
F(0))^{-1}[g] \Vert_\alpha\le K \Vert z\Vert_\alpha+C\Vert
g\Vert_\infty $, then $T$ maps $B(0,\delta)$ in itself, provided
that $\Vert g\Vert_\infty\le C^{-1}(1-K)\delta$.
\end{itemize}
Observe that the first of these two facts fixes the value of
$\delta$, whilst the last inequality holds choosing $\alpha$ -- free
so far -- small enough. This is one of the reasons why the constants
$M$ and $C$ must be independent on $\alpha$. Then $T$ admits a
unique fixed point on $B(0,\delta)$, which is the solution to our
problem \eqref{eq:inv} on the interval $[0,\alpha]$. In other words,
the function $\mathcal F$ is a local homeomorphism from
$B(0,\delta)\subseteq\mathcal X_\alpha$ to
$B(0,C^{-1}(1-K)\delta)\subseteq\mathcal Y_\alpha$.

To see that the solution is \emph{globally} unique on $\mathcal
X_\alpha$, let us argue as follows. Suppose $\bar z\in\mathcal
X_\alpha\setminus B(0,\delta)$ $(\bar z\ne z)$ solves the problem,
and let $\alpha_1\le\bar\alpha$ sufficiently small such that $\Vert
\bar z\Vert_{\alpha_1}\le\delta$. Then, observing $\Vert
g\Vert_{L^\infty([0,\alpha_1])}\le\Vert
g\Vert_{L^\infty([0,\alpha])}\le C^{-1}(1-K)\delta$, and recalling
that estimates \eqref{eq:F-lip}--\eqref{eq:Fprime} do not depend on
$\alpha$, one can argue as before to find that $\bar
z\vert_{[0,\alpha_1]}$ is the unique element of
$B(0,\delta)\subseteq\mathcal X_{\alpha_1}$ mapped into
$g\vert_{[0,\alpha_1]}\in B(0,C^{-1}(1-K)\delta)\subseteq\mathcal
Y_{\alpha_1}$. But, of course, $\Vert z\Vert_{\alpha_1}\le\Vert
z\Vert_\alpha<\delta$, so $z$ and $\bar z$ coincide on
$[0,\alpha_1]$, and therefore on all $[0,\alpha]$.

Therefore, to complete the proof of the existence and uniqueness for
the given Cauchy problem, just \eqref{eq:F-lip}--\eqref{eq:Fprime}
are to be proven. The first equation is a consequence of local
uniform Lipschitz continuity of $f_{,z}$. The second one needs some
more care: taken $\lambda\in\mathcal Y_\alpha$, we must consider the
Cauchy problem
\begin{equation}\label{eq:linCauchy}
h(s)\dot\xi(s)= \ell(s) \xi(s)+\lambda(s),\qquad\xi(0)=0,
\end{equation}
where $\ell(s):=\tfrac{\partial f}{\partial z}(s,0)$, that without
loss of generality we can suppose negative, $\forall s\in
[0,\beta]$. First, it is easily seen that \eqref{eq:linCauchy}
admits the unique solution $\xi\in\mathcal X_\alpha$,
\[
\xi(s)=e^{-\int_s^\alpha{\ell(t)}{h(t)^{-1}}\,\text
dt}\int_0^s\frac{\lambda(t)}{h(t)}
e^{\int_t^\alpha{\ell(\tau)}{h(\tau)^{-1}}\,\text d\tau}\,\text dt.
\]
Then $\Vert(\text d\mathcal
F(0))^{-1}[\lambda]\Vert_\alpha=\Vert\xi\Vert_\alpha\le(1+\ell_1)\Vert\xi
\Vert_\infty+\Vert\lambda\Vert_\infty$, where
$\ell_1=\Vert\ell\Vert_{L^\infty{[0,\beta]}}$. Moreover, called
$\ell_0=\sup_{[0,\beta]}\ell(s)<0$, it is easily seen that
$\Vert\xi\Vert_\infty\le-\tfrac1{\ell_0}\Vert\lambda\Vert_\infty$,
so it suffices to choose $C=1-\tfrac{\ell_1+1}{\ell_0}$, and
\eqref{eq:Fprime} is proven.

To prove last claim of the Theorem, let us suppose that $w(s)$ is a
function defined in $[0,\alpha]$ such that $h(s)\dot
w(s)=f(s,w(s))+g(s)$, and that $s_k$ is an infinitesimal and
monotonically decreasing sequence such that $w(s_k)\to 0$ as
$k\to\infty$. We want to prove that $w=z$, and therefore, it will
suffice to show that $\lim_{s\to 0} w(s)=0$.

First of all, observe that from the hypotheses, the equation
$f(s,z)+g(s)=0$ defines a continuous function
$\zeta(s):[0,\delta]\to\R$, such that $\zeta(0)=0$. In particular,
since $\frac{\partial f}{\partial z}(0,0)<0$, $\exists\rho>0$ such
that, in the rectangle $[0,\delta]\times[-\rho,\rho]$, it must be
$f(s,z)+g(s)<0$ (resp.: $>0$) if $z>\zeta(s)$ (resp.:$z<\zeta(s)$).

Let us now argue by contradiction, supposing the existence of an
infinitesimal sequence $\sigma_k$, that can be chosen with the
property $\sigma_k<s_k$, such that $\vert w(\sigma_k)\vert>\theta$
for some given constant $\theta$. Now, up to taking a smaller
constant $\delta$, then $\vert\zeta(s)\vert<\tfrac\theta 2,\,\forall
s\in[0,\delta]$. Then, for $k$ sufficiently large, $\vert
w(\sigma_k)\vert\ge\theta>\tfrac\theta
2\ge\sup_{[0,\delta]}\vert\zeta(s)\vert$. Therefore, $\forall
s<\sigma_k$, $\vert w(s)\vert>\vert w(\sigma_k)\vert$, which is a
contradiction since $w(s_k)\to 0$. Then $\lim_{s\to 0}w(s)=0$, and
the proof is complete.

\end{proof}

\begin{proof}[Proof of Lemma \ref{thm:perla}.]
Thanks to Proposition \ref{thm:div} we can carry on the proof
by studying qualitatively the solutions
of the ODE \eqref{eq:eva}. We will be interested in those solutions
which can be indefinitely prolonged on the right. With the variable
change $y=\sqrt{1+\rho}$, \eqref{eq:eva} becomes
\begin{equation}\label{eq:y}
\frac{\mathrm dy}{\mathrm d\phi}={\sqrt 3} (y^2-1)\left(u(\phi)y-1\right),
\end{equation}
where we recall that $u(\phi)$ is given by \eqref{eq:Y}.

Let us consider solutions of \eqref{eq:y} defined in
$[\phi_0,+\infty)$. If $\liminf_{\phi\to+\infty}{\vert
y(\phi)}{u(\phi)-1\vert}>0$ then, necessarily, $u(\phi)y(\phi)-1< 0$
(and $y(\phi)$ decreases, and goes to 1), otherwise the solution
would not be defined in a neighborhood of $+\infty$. In short, if
$y(\phi)$ is a solution defined in a neighborhood of $+\infty$, the
following behaviors are allowed:
\begin{enumerate}
\item\label{itm:1} either $y(\phi)$ is eventually weakly decreasing, and
tends to 1 (so that $\rho$ tends to 0),
\item\label{itm:2} or $\liminf_{\phi\to+\infty}{\vert
y(\phi)}{u(\phi)-1\vert}=0$.
\end{enumerate}
With the variable and functions changes
\[
s=e^{-{\sqrt 3}\phi},\qquad z=uy-1,
\]
equation \eqref{eq:y} takes the form
\begin{equation}\label{eq:z}
(s \,u)\dot z(s)=-(z+1)^2 z + (u^2+s \,\dot u) z+ s \,\dot u.
\end{equation}
But it can be easily seen, using \eqref{eq:uprime-infty}, and the
identity
$$s \,\dot u(s)=-\tfrac1{\sqrt 3}u'(\phi),$$ that \eqref{eq:z} satisfies
the assumption of Theorem \ref{thm:ode}, and so there exists a unique solution for the
Cauchy problem given by \eqref{eq:z} with the initial condition
$z(0)=0$, that is furthermore the only possible solution of the ODE
\eqref{eq:z} with $\liminf_{s\to 0}\vert z(s)\vert=0$, and this fact
results in a unique solution satisfying case \eqref{itm:2} above,
whereas  all other situations lead to case \eqref{itm:1}.
\end{proof}

Now, we are ready to prove the main result:

\begin{teo}\label{thm:sing-gen}
If $V(\phi)\in \mathfrak C$ then, except at most for a measure zero
set of initial data, there exists $t_s\in\R$ such that the scalar
field solution becomes singular at $t=t_s$, that is $\lim_{t\to
t_s^-}\epsilon(t)=+\infty,$ and $\lim_{t\to t_s^-}a(t)=0.$ Moreover
$\lim_{t\to t_s^-}\dot\phi(t)=+\infty.$
\end{teo}

\begin{proof}

We divide the proof into two parts, depending on the value of the
initial energy.

\noindent$\imath)$ \textbf{{case $\epsilon_0 > 2V_{*}.$}}

From \eqref{eq:energy} and \eqref{eq:efe2}--\eqref{eq:efe1}, it
easily follows that
\begin{equation}\label{eq:dotenergy}
\dot\epsilon(t)=2\sqrt 3\epsilon^{3/2}\frac1{1+\rho(t)}.
\end{equation}
But Lemma \ref{thm:perla} ensures that $\rho$ (defined in
\eqref{eq:rho}) goes to zero for almost every choice of initial
data. For this choice, $\dot\epsilon(t)>\epsilon^{3/2}$ in a left
neighborhood of $\sup\I$. Then $t_s < +\infty$ and $\lim_{t\to
t_s^-}\epsilon(t)=+\infty$ easily follows, using comparison theorems
in ODE. To prove that $\lim_{t\to t_s^-}\dot\phi(t)=+\infty$, it
suffices to observe that
\[
\frac{\epsilon(t)}{\dot\phi(t)^2}=\rho(t)+1\to 1.
\]
Finally, to show that $\lim_{t\to t_s^-}a(t)=0$, we first prove that
\begin{equation}\label{eq:inteps-as}
\lim_{t\to t_s^-}\int_0^{t}\sqrt\epsilon\,\mathrm d\tau=+\infty.
\end{equation}
Indeed, recalling that $\rho(t)$ is eventually bounded, $\exists
\bar t>0$ such that
$\epsilon(t)=\dot\phi^2(t)(1+\rho(t))\ge\kappa^2\dot\phi(t)^2$, for
some suitable constant $\kappa>0$, so
$\sqrt{\epsilon(t)}>\kappa\dot\phi(t)$, $\forall t>\bar t$, and then
\[
\int_{\bar t}^t\sqrt\epsilon\,\mathrm d\tau\ge\kappa\int_{\bar t}^t
\dot\phi(\tau)\,\mathrm
d\tau=\kappa(\phi(t)-\phi(\bar t)).
\]
The righthand side above diverges, so that  \eqref{eq:inteps-as} is
proved; but, using \eqref{eq:efe2}, we get
\[
\log\frac{a(t)}{a(0)}=\int_0^t\frac{\dot a}a\,\mathrm d\tau=-\frac1{\sqrt
3}\int_0^t\sqrt\epsilon\,\mathrm d\tau,
\]
from which $\lim_{t\to t_s^-}a(t)=0$ follows, since by the righthand
side is negatively divergent.

\noindent$\imath\imath)$ \textbf{{case $0 \leq \epsilon_0 \leq 2V_{*}$}}.

Suppose $\epsilon_{0} =0$. If the initial data are not at the zero level of $V$ (which consists only of a finite number of points
because of assumption (A1)), thanks to Lemma \ref{lem:eps0} in
Appendix \ref{sec:eps0}, where a result of local existence and uniqueness
for the solutions of equation \eqref{eq:efe1} satisfying
$\epsilon(0)=0$, but such that $\epsilon(t)>0$ for $t>0$,
we can assume $\epsilon_0 > 0$  without loss of generality.

Now suppose that $0 < \epsilon_{0} \leq \epsilon(t) \leq 2V_{*}$
$\forall t\in\I$ (otherwise the previously
obtained results would apply).We shall show that the solutions for
which this holds are non-generic.

To show this fact we observe that
in this case $\phi(t) \in [\alpha,\beta]$ for any $t \geq 0$ and, since $\dot
\phi$ is bounded, $\sup\I=+\infty$. Then Lemma \ref{lem:dotphi0} in Appendix \ref{sec:eps}
applies and we have

\[\lim_{t\to +\infty}\dot\phi(t)=0, \, \lim_{t\to +\infty}V'(\phi(t))=0.\]

Moreover by assumption (A1) there exists
$\phi_*$ critical point of $V$ such that $\lim_{t\to
+\infty}\phi(t)=\phi_*$, (with critical value $V(\phi_*) =
\lim_{t\to +\infty}\epsilon(t) \in ]0,2V_*]$.) Whenever $\phi_*$ is a
(nondegenerate) maximum point we can study the linearization of the
first order system equivalent to \eqref{eq:efe1} in a neighborhood
of the equilibrium point $(0,\phi_*)$ (such that $V(\phi_*) \geq 0$)
, as done in \cite{exp}, obtaining that there are only two solutions
$\phi_1$ and $\phi_2$ of \eqref{eq:efe1} such that $\lim\limits_{t
\rightarrow +\infty}\phi_i(t) = \phi_*$, $\lim\limits_{t \rightarrow
+\infty}\dot \phi_i(t) = 0$, $i=1,2$. Moreover there exists $t_* >
0$ such that $\phi_1(t) > \phi_*$ for all $t \geq t_*$ and
$\phi_2(t) < \phi_*$ for all $t \geq t_*$. It can be shown
\cite{exp} that,  if $\phi_*$ is a minimum point and $\phi$ starts
with initial data close to $(0,\phi_*)$, then $\phi$ moves far away
from $\phi_*$. Therefore, under the assumptions made, for almost
every choice of the initial data, the function $\phi(t)$ must be
such that its evolution cannot be contained in the bounded closed interval
$[\alpha,\beta]$,
namely, $\epsilon(t_1)\ge 2V_*$ for some $t_1>0$, which allows us to
apply the theory we already know to show that the singularity forms
for almost every choice of the initial data.
\end{proof}

\begin{rem}\label{rem:null_energy}
Since the potential can be negative the energy  $\epsilon$, could
in principle vanish "dynamically" (namely $\phi$ could be a
solution of $\dot\phi(t)^2=-2V(\phi(t)),\,\forall t$). However,
these functions are not solutions of Einstein field equations
\eqref{eq:original-efe1}--\eqref{eq:original-efe2}, since
$\epsilon(t)\equiv 0$ implies $a(t)=a_0$ and then
$\dot\phi(t)=0,\,V(\phi(t))=V(\phi_0)=0$. Thus, one is left with a
set of constant solutions with zero energy of the form
$(a_0,\phi_0)$, with $a_0$ positive constant and $V(\phi_0)=0$.
By the results obtained in Lemma \ref{lem:eps0} in appendix \ref{lem:eps0}
we therefore see that at the "boundary of the weak energy condition", local
uniqueness of the field equations may be violated if $V'(\phi_0)\ne
0$ and $V(\phi)=0$. However the set of the initial data for the expanding equations
intersecting such solutions in a finite time has zero measure (note
that the points $\phi_0$ such that $V(\phi_0)=0$ and $V'(\phi_0)\neq
0$ are isolated).
\end{rem}

\subsection{Examples}
\begin{ex}\label{ex:poly}
The above results hold for all potentials with polynomial leading
term at infinity (i.e. $\lambda^2\phi^{2n}$).
\begin{figure}
\begin{center}
\psfull \epsfig{file=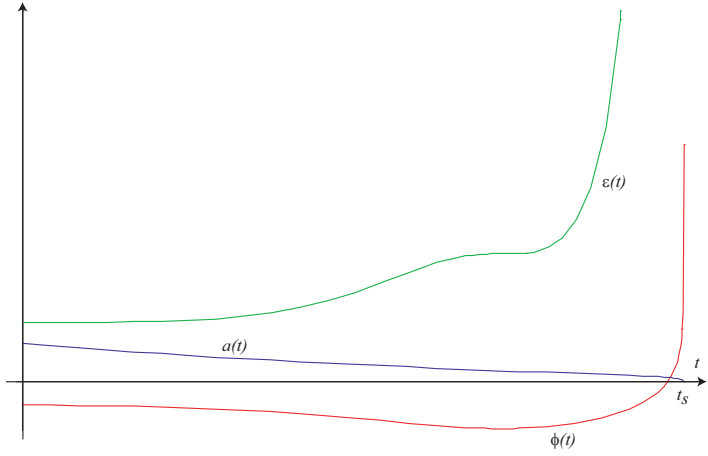, height=6cm} \caption{Behavior of the
function $\phi(t),\epsilon(t)$, and $a(t)$ with potential given by
$V(\phi)=1-\phi^2+\phi^4$. The initial conditions are
$\phi_0=-0.6,\,\dot\phi_0=0,\,a_0=1$. The time of collapse $t_s$
approximately equals $t_s=2.1$.}\label{fig:poly}
\end{center}
\end{figure}
For instance, for a quartic potential $V(\phi)=-\frac12
m^2\phi^2+\lambda^2\phi^4$, with $\lambda,m\ne 0$, the function
$u(\phi)$ goes as $\tfrac 2{\sqrt 3\phi}$ for $\phi\to +\infty$, and
all conditions listed above hold. The behavior for a particular
example from this class of potential is given in Figure
\ref{fig:poly}.
\end{ex}

\begin{ex}\label{ex:exp}
In the case of exponential potentials, the results hold for
asymptotic behaviors with leading term at infinity of the form $V_0
e^{2\sqrt{3} \lambda|\phi|}$ with $\lambda <1$. Consider for
instance $V(\phi)=V_0 \cosh(2\sqrt 3\lambda\phi)$, where
$V_0,\lambda>0$: the quantity $u(\phi)$ goes like $\lambda$, and so
\eqref{eq:u-infty} is verified if $\lambda<1$.
\begin{figure}
\begin{center}
\psfull \epsfig{file=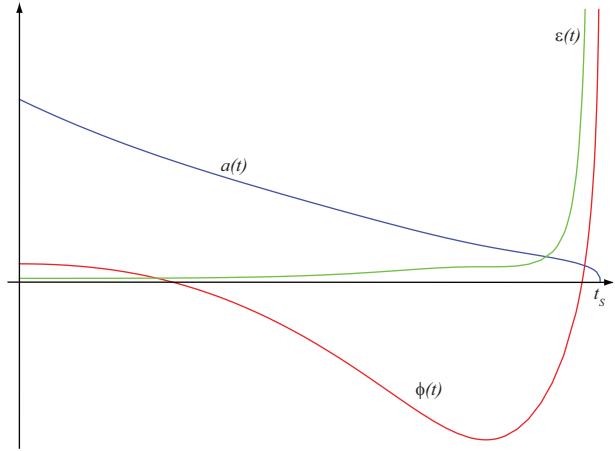, height=6cm} \caption{Same as
Figure \ref{fig:poly}, but now the potential is given by
$V(\phi)=\cosh(\sqrt 6\phi)/100$. The initial conditions are
$\phi_0=0.1,\,\dot\phi_0=0,\,a_0=1$. The time of collapse $t_s$
approximately equals $t_s=18.6$.}\label{fig:exp}
\end{center}
\end{figure}
See a particular situation from this class represented in Figure
\ref{fig:exp}.

\end{ex}

\begin{ex}\label{ex:decay}
Decaying exponential potentials can be considered as well. For
instance for $V(\phi)=(1-e^{-\alpha\sqrt{\phi^2+\gamma^2}})^2$ the
function $u(\phi)$ behaves like $e^{-\alpha|\phi|}$, so it goes to
zero at $\pm\infty$.
\begin{figure}
\begin{center}
\psfull \epsfig{file=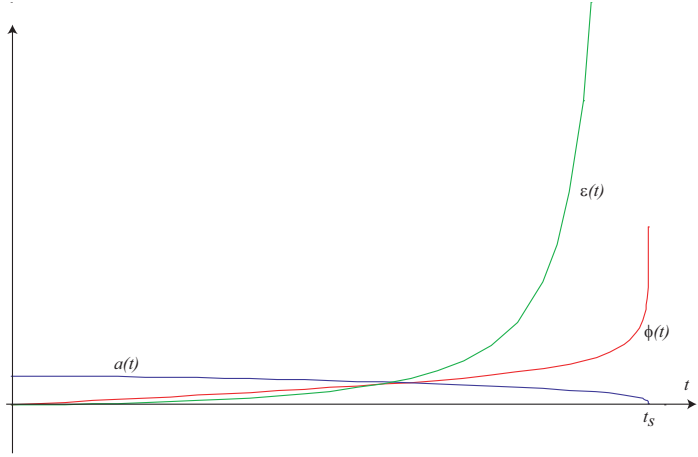, height=6cm} \caption{Same as Figure
\ref{fig:poly}, where now $V(\phi)=2(1-e^{-\sqrt{x^2+1}})^2-\tfrac16
e^{-(x+4)^2}-\tfrac12 e^{-(x-4)^2}+1$. The initial conditions are
$\phi_0=0,\,\dot\phi_0=\sqrt{-2V(0)},\,a_0=1$. The time of collapse
$t_s$ approximately equals $t_s=1.8$.}\label{fig:decay}
\end{center}
\end{figure}
Figure \ref{fig:decay} represent a situation where some corrections
terms have been added in order to obtain a potential with more than
one critical points. Even choosing initial data such that
$\epsilon_0=0$, the solution diverges in a finite amount of time, as
showed in general in the second part of the proof of Theorem \ref{thm:sing-gen}.
\end{ex}

\section{Gravitational collapse models}

In what follows, we construct models of collapsing objects composed
by homogeneous scalar fields. To achieve this goal we must match a
collapsing solution (considered as the interior solution in matter)
with an exterior spacetime. The natural choice for the exterior is
the so--called \emph{generalized Vaidya solution}
\begin{equation}\label{eq:Va}
\text ds_{\mathrm{ext}}^2=-\left(1-\frac{2M(U,Y)}Y\right)\,\mathrm
dU^2-2\,\mathrm
dY\,\mathrm dU + Y^2\,\mathrm d\Omega^2,
\end{equation}
where $M$ is an arbitrary (positive) function (we refer to \cite{ww}
for a detailed physical discussion of this spacetime, which is
essentially the spacetime generated by a radiating object). The
matching is performed along a hypersurface $\Sigma$ which, in terms
of a spherical system of coordinates for the interior, has the
simple form  $r=r_b=const$. The Israel junction conditions at the
matching hypersurface read as follows \cite{hsf}:
\begin{align}
&M(U(t),Y(t))=\frac12 r_b^2 a(t)\dot a(t)^2,\label{eq:mass-match1}\\
&\frac{\partial M}{\partial Y}(U(t),Y(t))=\frac12 r_b^3(\dot a(t)^2+2
a(t)\ddot a(t)),\label{eq:mass-match2}
\end{align}
where the functions $(Y(t),U(t))$ satisfy
\begin{equation}\label{eq:cond}
Y(t)=r_b a(t),\qquad\frac{\text d U}{\text d
t}(t)=\frac1{1+\dot a(t) r_b}.
\end{equation}
The two equations \eqref{eq:mass-match1}--\eqref{eq:mass-match2} are
equivalent to require that Misner--Sharp mass $M$ is continuous and
$\tfrac{\partial M}{\partial U}=0$ on the junction hypersurface
$\Sigma$ (using \eqref{eq:mass-match1}--\eqref{eq:cond} together
with \eqref{eq:original-efe1}--\eqref{eq:Bianchi}, it can be seen
that in this way also the equation of motion for the scalar field
remains smooth on the matching hypersurface).

The endstate of the collapse of these ``homogeneous scalar field
stars'' is analyzed in the following theorem.

\begin{teo}\label{thm:endstate}
Except at most for a measure zero set of initial data, a scalar
field "star" collapses to a black hole.
\end{teo}
\begin{proof}

It is  easy to check that the equation of the apparent horizon for
the metric \eqref{eq:g} is given by $r^2\dot a(t)^2=1$. If $\dot a$
is bounded, one can choose the junction surface $r=r_b$ sufficiently
small such that $\dot a^2(t)<\tfrac1{r^2}$, $\forall
(t,r)\in[0,t_s]\times[0,r_b]$, and so $\left(1-\tfrac{2M}R\right)$
is bounded away from zero near the singularity. As a consequence,
one can find in the exterior portion of the spacetime \eqref{eq:Va},
null radial geodesics which meet the singularity in the past, and
therefore the singularity is naked; otherwise, if $\dot a^2$ is
unbounded, the trapped region forms and the collapse ends into a
black hole \cite{hsf}. Now, we have proved in Lemma \ref{thm:perla}
the existence of $\rho_\infty=\lim_{\phi\to\infty}\rho(\phi)$ (which
is also equal to $\lim_{t\to t_s^-}\rho(t)$), and that, actually,
$\rho_\infty$ vanishes except for a zero--measured set of initial
data. On the other hand, using \eqref{eq:original-efe2}, we get
\begin{equation}\label{eq:efe-nu2}
\ddot a=-\frac{\dot a^2}a\left(\frac{2-\rho}{1+\rho}\right),
\end{equation}
and therefore $\ddot a\le-\tfrac{\dot a^2}a$ eventually holds in a
left neighborhood of $t_s$, say $[t_0,t_s[$, where $\dot a$ is
decreasing. It follows that $\forall t\in]t_0,t_s[$, $\dot a(t)\le
\dot a(t_0)$, which is negative by hypotheses. Then
\[
\dot a(t)-\dot a(t_0)
=\int_{t_0}^t\ddot a(\tau)\,\mathrm d\tau\le-\int_{t_0}^t\frac{\dot
a(\tau)^2}{a(\tau)}\,\mathrm d\tau\le -\dot a(t_0)\int_{t_0}^t\frac{\dot
a(\tau)}{a(\tau)}\,\mathrm d\tau,
\]
that diverges to $-\infty$. Then $\dot a(t)$ is unbounded.
\end{proof}

\section{Naked singularities as examples of non-generic situations}\label{rem:last}

As we have seen, the formation of naked singularities in the above
discussed models is {\it generically } forbidden by theorem
\ref{thm:endstate}. Thus, the theorem does not exclude the existence
of initial data which give rise to naked singularities for the given
potential, it only assures that such data are non-generic. Actually,
there are examples of such solutions containing naked singularities
in the literature, and it is our aim here to show how they fit in
this scenario.

To discuss this issue, it is convenient to reformulate the dynamics
in the following form \cite{hsf}. Since $a(t)$ is strictly
decreasing in the collapsing case, we re-write equations
\eqref{eq:energy}, \eqref{eq:efe2} and \eqref{eq:efe1} using $a$
instead of $t$ as the independent variable. Setting
\begin{equation}\label{eq:ea}
\epsilon (a)=3\left(\frac{\psi(a)}a\right)^2,
\end{equation}
we get:
\begin{subequations}
\begin{align}
\dot a&=-\psi(a),\label{e1}\\
\left(\frac{\text d\phi}{\text
da}\right)^2&=\frac1{a^2}\left(1-a\frac{\psi'(a)}{\psi(a)}\right),\label{e2}\\
V&=\left(\frac{\psi(a)}a\right)^2\left(1+\frac12a\frac{\psi'(a)}{\psi(a)}\right),\label{e3}
\end{align}
\end{subequations}
In the above equations, $V(\phi)$ plays of course the same role of
"equation of state" as before. However, one can try to prescribe a
part of the dynamics and then search if there is a choice of the
potential and a set of data which assure that this dynamics is
actually an admissible one for the system at hand (of course, in
this way, only the "on shell" value of the potential can be
reconstructed). In particular, a somewhat natural approach can be
that of prescribing the energy ($\epsilon (a)$ or equivalently $\Psi
(a)$).

In \cite{joshi}, the case in which the energy density behaves like
$a^{-\nu}$ with $\nu$ positive constant is considered. It is then
shown that for $\nu<2$ the system forms a naked singularity, and the
existence of an interval of values of $\nu$ is interpreted as a {\it
generic} violation of cosmic censorship (that is anyway restored
when loop quantum gravity modifications are taken into account, see
\cite{joshiprl}). However, due to the equation  \eqref{e3} above,
the {\it on shell} functional form of the potential for this example
can be calculated to be $V(\phi)=V_0 e^{2\sqrt
3\lambda\vert\phi\vert}$ with $\lambda=\sqrt{\nu /6}$. This
potential belongs to the class which has been treated in our
previous example \ref{ex:exp}, and therefore we know that a generic
choice of data must give rise to a blackhole (for this class the
function $u(\phi)=V'/(2\sqrt 3 V)$ behaves like a \emph{strictly
positive} constant $\lambda<1$). However, as we know from the proof
of Lemma \ref{thm:perla}, a non generic situation can occur, when
$\lim_{t\to t_s^-}u\sqrt{1+\rho}=1$ (where $\rho=2V/\dot\phi^2$);
which means that
$\lim_{\phi\to\infty}\rho(\phi)=\tfrac1{\lambda^2}-1=6/\nu-1$.
Evaluating the function $\rho$ for the solution under study, we get
{\it exactly }$\rho=6/\nu-1$. Therefore, the case discussed in
\cite{joshi,joshiprl} is precisely the non-generic one. The
constraint on the data which implies non-genericity comes, of
course, from the fact that imposing a certain behavior on the
function $\psi (a)$ selects a two-dimensional set in the space of
the data, characterized by the algebraic constraint
$\dot\phi(0)^2+2V(\phi(0))=3(\psi(a(0))/a(0))^2$.

To complete the analysis, we observe that, although the evolution of
non generic data is not predicted by Theorem \ref{thm:sing-gen}, we
can easily obtain information also in this case. To this aim,
observe that, since $\rho$ is finite, equation \eqref{eq:dotenergy}
can be used to prove singularity formation in a finite amount of
comoving time with an argument similar to that used in the proof of
Theorem \ref{thm:sing-gen}. Moreover, the metric function $a(t)$
behaves asymptotically as $a_0(t_s-t)^{1/(3\lambda^2)}$: then, if
$\lambda<\sqrt 3/3$, the function $\dot a(t)$ is bounded near the
singularity, confirming naked singularity formation (this choice of
$\lambda$ corresponds exactly to the choice of $\nu<2$ in
\cite{joshi}).

\begin{figure}
\begin{center}
\psfull \epsfig{file=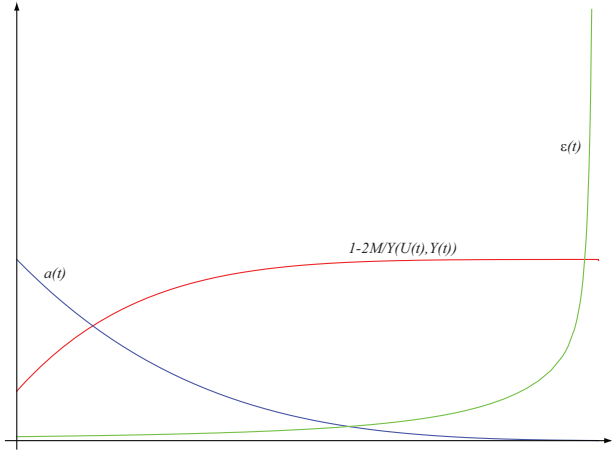, height=6cm}
\caption{An example of naked singularity for $V(\phi)=0.01 e^\phi$ for $\phi\ge 0$. The initial conditions are
$\phi_0=0,\,\dot\phi_0=550^{-1/2},\,a_0=1$. The matching with external solution is performed at $r_b=100$. The time of collapse
$t_s$ approximately equals $t_s=38.8$.}\label{fig:exp-naked}
\end{center}
\end{figure}

Figure \ref{fig:exp-naked} sketches an example of this kind:
the initial data are chosen in the non generical way such that
$\rho$ goes like a positive constant as $t\to t_s$.
The singularity forms (i.e. $\epsilon(t)$ diverges in a finite time),
but choosing $r_b$ not too large, the function $1-2M/Y$ on the junction hypersurface $r=r_b$
remains strictly positive along the evolution up to the singularity.

\bigskip

The above example suggests that hypotheses on the asymptotic
behavior of the function $u(\phi)$ can be added, to obtain
singularity formation also in the non generic case: indeed, if the
potential $V(\phi)$ satisfies the condition
\begin{equation}\label{eq:nongen}
\liminf_{\phi\to\pm\infty}|u(\phi)|> 0,
\end{equation}
i.e. $u(\phi)$ is definitely bounded away from zero, then $\rho$ is
finite also in the non generic case, and equation
\eqref{eq:dotenergy} in the proof of Theorem \ref{thm:sing-gen} can
be used to prove the formation of the singularity in a finite amount
of time. To analyze the nature of this singularity observe that, if
\begin{equation}\label{eq:nongen-bh}
\liminf_{\phi\to\pm\infty}|u(\phi)|>\tfrac{\sqrt 3}3,
\end{equation}
the argument in the proof of Theorem \ref{thm:endstate} applies, and
therefore the collapse ends into a black hole. On the other side, if
\begin{equation}\label{eq:nongen-ns}
\limsup_{\phi\to\pm\infty}|u(\phi)|<\tfrac{\sqrt 3}3,
\end{equation}
then $\ddot a$ is eventually positive, and this implies that $\dot
a$ is bounded, so that the formation of the apparent horizon is
forbidden. As a consequence, the collapse ends into a naked
singularity.

\section{Discussion and Conclusions}

We have discussed here the qualitative behavior of the solutions of
the Einstein field equations with homogeneous scalar fields sources
in dependence of the choice of the self-interacting potential. It
turns out that whenever a potential satisfies a certain set of
general conditions singularity formation occurs for almost every
choice of initial data. Matching these singular solutions with a
Vaidya "radiating star" exterior we obtained models of gravitational
collapse which can be viewed as the scalar-field generalization of
the Oppenheimer-Snyder collapse model, in which a dust homogeneous
universe is matched with a Schwarzschild solution (the Schwarzschild
solution can actually be seen as a special case of Vaidya).

The Oppenheimer-Snyder model, as is well known, describes the
formation of a covered singularity, i.e. a blackhole; it is actually
the first and simplest model of blackhole formation ever discovered.
The same occurs here: indeed we show that homogeneous scalar field
collapse generically forms a blackhole. The examples of naked
singularities which were found in recent papers \cite{hsf,joshi},
turn out to correspond to very special cases which, mathematically,
are not generic. Therefore, our results here support the weak cosmic
censorship conjecture, showing that naked singularities are not
generic in homogeneous, self-interacting scalar field collapse.

Non-genericity was already well known for \emph{non
self-interacting} (i.e. $V(\phi)=0$) spherically symmetric scalar
fields. Whether the results obtained here can actually be shown to
hold also in the much more difficult case of both inhomogeneous and
self-interacting scalar fields remains an open problem.

\appendix

\section{Some properties of global solutions}\label{sec:eps}

We give here the proof that velocity and acceleration of solutions
which extend indefinitely in the future must vanish asymptotically,
if the values of the derivatives of the potential remain bounded on
the corresponding flow.

\begin{lem}\label{lem:dotphi0}
Let $\phi(t)$ a finite-energy solution of \eqref{eq:efe1}, with
$V\in\mathfrak C$. Then:

1) $\phi$ can be extended for all $t>0$,

2) if $V'$ is bounded on the flow, then $\lim_{t\to
+\infty}\dot\phi(t)=0$,

3) if also $V''$ is bounded on the flow then $\lim_{t\to
+\infty}\ddot\phi(t)=0$.

\end{lem}

\begin{proof}
Since $\epsilon$ is bounded and $V$ is bounded from below,
$\dot\phi$ is bounded as well, and thus $\phi$  can be extended for
all $t$ (actually the existence of a lower bound for $V$ is the
unique condition among A1)-A3) used here and below).

Using \eqref{eq:3} we have
\[
\int_0^{+\infty}\dot\phi(t)^2\,\mathrm
dt=\int_0^\infty\frac{\dot\epsilon(t)}{2\sqrt
3\sqrt{\epsilon(t)}}\,\mathrm dt<+\infty,
\]
so there exists a sequence $t_k\to+\infty$ such that
$\dot\phi(t_k)\to 0$. By contradiction, suppose that
$\exists\rho>0$, and a sequence $s_k\to t_k$ -- that can be taken
such that $s_k>t_k$ -- with $\vert\dot\phi(s_k)\vert>\rho$. Let
$\bar k$ such that $\vert\dot\phi(t_k)\vert<\tfrac\rho 2,\,\forall
k\ge \bar k$, and let $\tau_k,\sigma_k$ sequences such that $t_k\le
\tau_k<\sigma_k\le s_k$, and
\[
|\dot\phi(\tau_k)|=\frac\rho 2,\qquad
|\dot\phi(\sigma_k)|=\rho,\qquad\frac\rho
2\le\dot\phi([\tau_k,\sigma_k])|\le\rho.
\]
Note that $\epsilon$ bounded implies $\dot \phi$ bounded, while by
assumption, $V'(\phi(t))$ is bounded. Therefore by
\eqref{eq:initialefe1}, there exists $M > 0$ such that
\[ |\ddot\phi(t)| \leq M \qquad \forall t.\]
Therefore
\[
\frac\rho
2\leq|\dot\phi(\sigma_k)-\dot\phi(\tau_k)|\leq\int_{\tau_k}^{\sigma_k}|\ddot
\phi(t)|\,\mathrm dt\le M(\sigma_k-\tau_k),
\]
that is $\sigma_k-\tau_k\ge\tfrac\rho{2M}$, and therefore
\[
\int_{0}^{+\infty}\dot\phi(t)^2\,\mathrm
dt\ge\sum_k\int_{\tau_k}^{\sigma_k}\dot\phi(t)^2\,\mathrm
dt\ge\sum_k(\sigma_k-\tau_k)\cdot\left(\frac\rho
2\right)^2\ge\sum_k\frac{\rho^3}{8M},
\]
that diverges. This is a contradiction, and therefore it must be
$\lim_{t\to+\infty}\dot\phi(t)=0$.

To prove that the acceleration also vanish, let us first observe that
$\exists t_k\to+\infty$ such that $V'(\phi(t_k))\to 0$ -- otherwise,
there would exists $\kappa>0$ such that $\vert
V'(\phi(t))\vert\ge\kappa$ definitely, which would imply, in view of
\eqref{eq:initialefe1}, that $\vert\ddot\phi(t)\vert\ge\kappa/2$,
that is absurd since $\lim_{t\to+\infty}\dot\phi(t)=0$.

Then, let us suppose by contradiction the existence of a constant
$\rho>0$, and a sequence $s_k\to+\infty$ such that $t_k<s_k$ and
$\vert V'(\phi(s_k))\vert\ge\rho$. Therefore, one can choose
$\sigma_k,\tau_k$ sequences such that $t_k\le\tau_k<\sigma_k\le
s_k$, and
\[
\vert V'(\phi(\tau_k))\vert=\frac\rho 2,\qquad \vert
V'(\phi(\sigma_k))\vert=\rho,\qquad \frac\rho 2\le\vert
V'(\phi(t))\vert\le \rho,\,\forall t\in [\tau_k,\sigma_k].
\]
Then, since by assumption $V''(\phi(t))$ is bounded, there exists a
constant $L>0$ such that
\[
\frac\rho 2=\vert
V'(\phi(\sigma_k))-V'(\phi(\tau_k))\vert\le\int_{\tau_k}^{\sigma_k}\vert
V''(\phi(t))\vert\,\vert\dot\phi(t)\vert\,\text dt\le L
(\sigma_k-\tau_k).
\]
But for sufficiently large $k$ let us observe that
\eqref{eq:initialefe1} implies $\vert\ddot\phi(t)\vert\ge\rho/4$,
$\forall t\in [\tau_k,\sigma_k]$, and therefore
$\vert\dot\phi(\sigma_k)-\dot\phi(\tau_k)\vert=\vert\int_{\tau_k}^{\sigma_k}
\ddot\phi(t)\text dt\vert\ge\frac{\rho^2}{8L}$ that is a
contradiction since $\dot\phi\to 0$. Thus $\lim_{t\to
+\infty}V'(\phi(t))=0$ vanishes, and equation \eqref{eq:initialefe1}
implies that also $\ddot\phi$ vanishes in the same limit.

\end{proof}

\section{Local existence/uniqueness of solutions with initial
zero--energy}\label{sec:eps0}

\begin{lem}\label{lem:eps0}
Let $\phi_0,\,v_0$ such that $v_0^2+2V(\phi_0)=0$. Then, there
exists $t_*>0$ such that the Cauchy problem
\begin{equation}\label{eq:eps0}
\begin{cases}
&\ddot\phi(t)=-V'(\phi(t))+\sqrt{3(\dot\phi(t)^2+2V(\phi(t)))}\,\dot\phi(t),
\\
&\phi(0)=\phi_0,\\
&\dot\phi_0=v_0,
\end{cases}
\end{equation}
has a unique solution $\phi(t)$ defined in $[0,t_*]$ with the
property
\begin{equation}\label{selezione}
\epsilon(t)=3\left(\int_0^t\dot\phi(s)^2\,\text ds\right)^2,\,
\forall t\in]0,t_*].
\end{equation}

Moreover if $(\phi_{0,m},v_{0,m}) \rightarrow (\phi_0,v_0)$,
$(v_{0,m})^2 + 2V(\phi_{0,m}) = 0$ and $\phi_m$ is the solution of
\eqref{eq:eps0} with initial data $(\phi_{0,m},v_{0,m})$ satisfying
condition \eqref{selezione}, it is $\phi_m \rightarrow \phi$ with
respect to the $C^2$-norm in the interval $[0,t_*]$.
\end{lem}
\begin{proof}
Let us consider the "penalized" problem
\begin{equation}\label{eq:pen}
\begin{cases}
&\ddot\phi(t)=-V'(\phi(t))+\sqrt{3(\dot\phi(t)^2+2V(\phi(t))+\frac1{n^2})}\,
\dot\phi(t),\\
&\phi(0)=\phi_0,\\
&\dot\phi_0=v_0,
\end{cases}
\end{equation}
that has a unique local solution $\phi_n$. If $\phi_n$ is not
defined $\forall t\ge 0$, let $\I_n$ be the set
\[
\I_n=\{t\in\R\,:\,|\phi_n(s)| \leq |\phi_0|
+1,\,|\dot\phi_n(s)|\le|v_0|+1,\,\forall s\ge t\}.
\]
Of course, $\I_n\ne\emptyset$ and, called $t_n=\sup\I_n$, if $t_n$
is finite, then $|\dot\phi_n(t_n)|=|v_0|+1$, or $|\phi_n(t_n)| =
|\phi_0| +1$. Now assume $|\phi_n(t_n)| = |\phi_0| +1$. Then
\[
1 = \vert \phi_n(t_n)-\phi_0\vert\le
\int_0^{t_n}\vert\dot\phi_n(s)\vert\,\text ds\le(|v_0|+1)t_n.
\]
Analogously if  $|\dot\phi_n(t_n)|=|v_0|+1$ we have
\[
1 = \vert \dot \phi_n(t_n)- v_0\vert\le
\int_0^{t_n}\vert\ddot\phi_n(s)\vert\,\text ds.
\]
Since $|\phi_n(t)| \leq |\phi_0| +1$ and $|\dot\phi_n(s)|\le|v_0|+1$
for all $t \in [0,t_n]$, and $\phi_n$ solves \eqref{eq:pen}, we see
that there exists $C$ independent of $n$ such that $\vert\ddot
\phi(t)\vert \leq C$ for all $t \in [0,t_n]$. Therefore in this
second case we obtain $1 \leq Ct_n$.

Then $t_*:=\inf_n t_n>0$ (we set $t_*=1$ if $t_n=+\infty$ $\forall
n$). Moreover $\vert\ddot \phi_n\vert$ is uniformly bounded in
$[0,t_*]$, then up to subsequences, there exists a $\mathcal C^1$
function $\phi(t)$, solution of \eqref{eq:eps0}, such that
$\phi_n\to\phi$ and $\dot\phi_n\to\dot\phi$ uniformly on $[0,t_*]$.

Now, consider
$\epsilon_n(t):=\dot\phi_n(t)^2+2V(\phi_n(t))+\tfrac1{n^2}$. We have
\begin{equation}\label{energiaappr}
\dot\epsilon_n(t)=2\sqrt 3\sqrt{\epsilon_n(t)}\dot\phi_n(t)^2.
\end{equation}
Then $\epsilon$ is not decreasing, while $\epsilon(0) = \frac1n$.
Then is uniformly bounded away from zero and therefore by
\eqref{energiaappr}, dividing by $\sqrt{\epsilon_n}$ and integrating
gives $\sqrt{\epsilon_n(t)}=\tfrac1{n^2}+\sqrt 3\int_0^t
\dot\phi_n(s)^2\,\text ds$. Therefore passing to the limit in $n$ we
obtain $\epsilon(t)=\dot\phi(t)^2+2V(\phi(t))=3\left(\int_0^t
\dot\phi(s)^2\,\text ds\right)^2$ for all $t \in [0,t_*]$ obtaining
the proof of the existence of a solution.

The uniqueness of such a solution can be obtained by a contradiction
argument. Assuming $\phi$ and $\psi$ solutions, and called
$\theta=\phi-\psi$, one can obtain, using \eqref{eq:eps0}, the
estimate
\[
\vert\dot\theta(t)\vert\le K_1\int_0^t\vert\theta(s)\vert\,\text ds+
K_2\int_0^t\vert\dot\theta(s)\vert\,\text ds,
\]
for suitable constants $K_1,K_2$. Setting
$\rho(t)=\vert\theta(t)\vert+\vert\dot\theta(t)\vert$, and observing
that $\rho(0)=0$, it is not hard to get the estimate $\rho(t)\le
(K_1+K_2+1)\int_0^t\rho(s)\,\text ds$, and then $\rho\equiv 0$ from
Gronwall's inequality.

Finally using Gronwall's Lemma as above we obtain also the
continuity with respect to the initial data.
\end{proof}

\begin{rem}\label{rem:eps0rev}
Reversing time direction in the above discussed problem
\eqref{eq:eps0} yields a results of genericity for expanding
solutions such that the energy $\epsilon(t)$ vanishes at some finite
time $T$.
\end{rem}

\end{document}